\documentclass[aps,pra,superscriptaddress,twocolumn,showpacs,floatfix]{revtex4-2}

\usepackage{amssymb,amsmath,amsfonts,mathrsfs,bm, bbm, graphicx, epsfig, epstopdf}
\usepackage[version=3]{mhchem}
\usepackage{braket}
\usepackage{hyperref}
\usepackage{color}
\usepackage{adjustbox}
\usepackage{csquotes}
\usepackage{float}
\usepackage{colortbl}
\usepackage{siunitx}
\DeclareSIUnit{\nothing}{\relax}
\usepackage{makecell}
\usepackage[T1]{fontenc}

\date{\today}
\begin{document}
\title{Quantum Optical Memory for Entanglement Distribution}

\author{Yisheng Lei}
\thanks{These two authors contributed equally}
\affiliation{Birck Nanotechnology Center and Purdue Quantum Science and Engineering Institute, Elmore Family School of Electrical and Computer Engineering, Purdue University, West Lafayette, Indiana 47907, USA}
\affiliation{Department of Physics and Astronomy, Purdue University, West Lafayette, Indiana 47907, USA}

\author{Faezeh Kimiaee Asadi}
\thanks{These two authors contributed equally}
\affiliation{Institute for Quantum Science and Technology, and Department of Physics \& Astronomy, University of Calgary, 2500 University Drive NW, Calgary, Alberta T2N 1N4, Canada}

\author{Tian Zhong}
\affiliation{Pritzker School of Molecular Engineering, University of Chicago, Chicago, Illinois 60637, USA}

\author{Alexander Kuzmich}
\affiliation{Department of Physics, University of Michigan, Ann Arbor, Michigan 48109, USA}

\author{Christoph Simon}
\affiliation{Institute for Quantum Science and Technology, and Department of Physics \& Astronomy, University of Calgary, 2500 University Drive NW, Calgary, Alberta T2N 1N4, Canada}

\author{Mahdi Hosseini}
\thanks{Corresponding author}
\email{mh@purdue.edu}
\affiliation{Birck Nanotechnology Center and Purdue Quantum Science and Engineering Institute, Elmore Family School of Electrical and Computer Engineering, Purdue University, West Lafayette, Indiana 47907, USA}
\affiliation{Department of Physics and Astronomy, Purdue University, West Lafayette, Indiana 47907, USA}

\begin{abstract}
Optical photons are powerful carriers of quantum information, which can be delivered in free space by satellites or in fibers on the ground over long distances. Entanglement of quantum states over long distances can empower quantum computing, quantum communications, and quantum sensing. Quantum optical memories can effectively store and manipulate quantum states, which makes them indispensable elements in future long-distance quantum networks. Over the past two decades, quantum optical memories with high fidelity, high efficiencies, long storage times, and promising multiplexing capabilities have been developed, especially at the single photon level. In this review, we introduce the working principles of commonly used quantum memory protocols and summarize the recent advances in quantum memory demonstrations. We also offer a vision for future quantum optical memory devices that may enable entanglement distribution over long distances.  
\end{abstract}

\maketitle{}

\section{Introduction}
\begin{center}
\begin{figure}[!ht]
	\centerline{\includegraphics[width=1\columnwidth, angle=0]{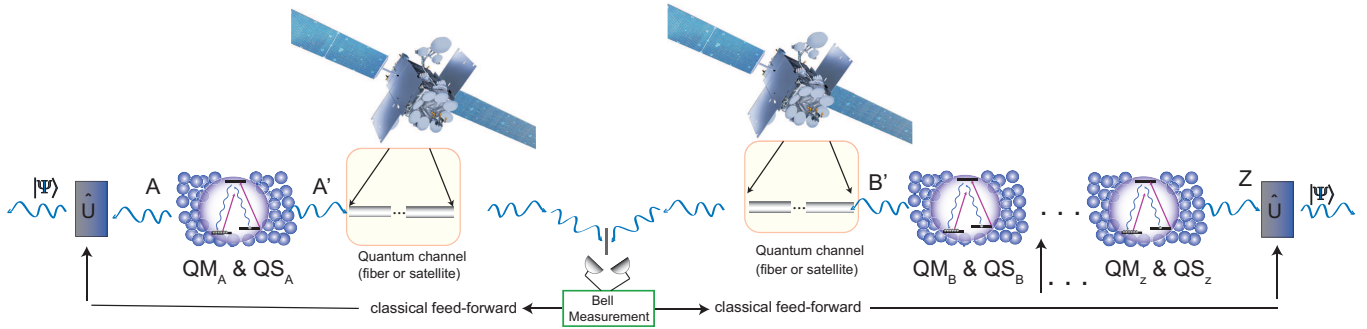}}
	\caption{
	Illustration of a quantum memory application in optical entanglement distribution. A long communication channel is divided into multiple links with quantum memories (QM) and quantum sources (QS) at their ends. Joint measurement (or Bell-state measurement) of photons from neighboring quantum memories projects the state of memories into an entangled state.  Repeating the measurement until the first and last memories are entangled by multiple entanglement swapping operations results in entanglement distribution between memories A and Z without a long-distance quantum channel connecting the two nodes. The quantum channels connecting the memories can be fiber or satellite links.}
	\label{fig:fig1_repeater}
\end{figure}
\end{center}

Quantum optical states such as entangled photons can carry quantum information enabling distribution of quantum information in a network setting \cite{kimble2008quantum, wehner2018quantum, simon2017towards, lloyd2004infrastructure}. However, without quantum memories, the fiber-based communication distance is limited to ~100km due to losses in fibers, which exponentially degrades the degree of quantum correlations. For this reason, quantum memories were proposed as part of a quantum repeater \cite{briegel1998quantum, duan2001long, sangouard2011quantum, muralidharan2014ultrafast, muralidharan2016optimal, rozpkedek2021quantum, fukui2021all}  to extend the communication distance. Quantum memories storing entangled photons can themselves be entangled. As seen in Fig. \ref{fig:fig1_repeater}, entanglement between neighboring memories can be established by Bell state (or joint) measurement on nearby initially unentangled memories. Subsequent measurement eventually leads to the creation of entanglement between memories in distant nodes. The information can then be teleported from node A to Z without directly sending quantum states, $\ket{\psi}$, through the channel \cite{bennett1993teleporting, bouwmeester1997experimental}. A quantum network of this kind has applications in distributed quantum computing, global parameter estimation using entangled sensors, and secure communication \cite{wehner2018quantum, komar2014quantum, gottesman2012longer}. Other applications of quantum optical memories include deterministic single photon  \cite{kaneda2019high, dideriksen2021room} or multi-photon  \cite{kaneda2017quantum, chrapkiewicz2017high} generation, coherent optical-to-microwave transduction \cite{han2021microwave}, quantum imaging \cite{mazelanik2021real}, enhanced quantum sensing \cite{zaiser2016enhancing}, non-destructive photon detection \cite{hosseini2016partially, niemietz2021nondestructive}, linear optical quantum computing \cite{knill2001scheme} and fundamental science in space \cite{mol2023quantum}.

In this review, we discuss the basic concepts of light storage as applied to atomic quantum memories and review recent advances in quantum memory development. 

\section{Tutorial}

\subsection{Basic Concepts}
\begin{center}
\begin{figure}[!ht]
	\centerline{\includegraphics[width=1\columnwidth, angle=0]{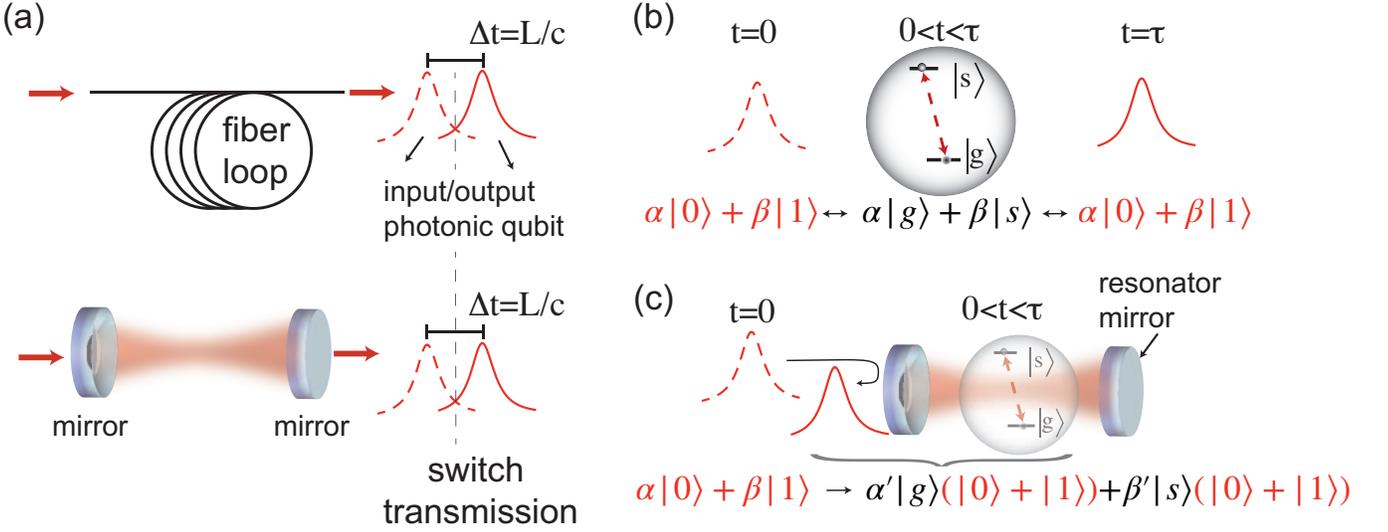}}
	\caption{
	Forms of quantum memories. (a) A fiber loop (top) or a pair of mirrors (bottom) can be used to delay optical qubits by controlling the propagation length $L$. (b) A single or ensemble of atoms can be controlled to coherently and reversibly map quantum information to/from a photon from/to atomic coherence.  (c) An atom inside a low-loss optical resonator can enhance light-atom interactions to the point that reflecting a single photon off the cavity mirror can result in the entanglement between the photon and the atomic spin.}
	\label{fig:fig2_QMconcept}
\end{figure}
\end{center}

In the simplest form, a quantum optical memory can be realized by passive delay lines such as a fiber loop or a pair of mirrors increasing the propagation length, $L$ (see Fig. \ref{fig:fig2_QMconcept}(a)).  To implement an atomic memory, on the other hand, quantum fluctuation from an optical field should be coherently mapped to stationary excitations of atomic spins\cite{duan2010colloquium} (see Fig. \ref{fig:fig2_QMconcept}(b)). The mapping should be ideally reversible and on-demand, and the process should not add noise to the retrieved optical state. Consider a simple optical qubit in a superposition of two number states $\alpha\ket{0}$ and $\alpha\ket{1}$ written as: $\alpha\ket{0}+\beta e^{i\theta}\ket{1}$. In practice, other types of qubit encodings, e.g. time-bin or polarization qubits, are used for long-distance communication. To store information in an atomic memory, the amplitude and phase of the qubit should be transferred to coherence between two energy levels of an atom, e.g. $\ket{g}$ and $\ket{s}$, written as: $\alpha\ket{g}+\beta e^{i\theta}\ket{s}$. The process does not necessarily store the energy of the photon but just the information encoded in $\alpha / \beta$ and $\theta$. In atomic systems, the storage essentially corresponds to a coherent and reversible mapping of flying electromagnetic excitations into stationary qubits. 

We first classify quantum optical memories (QOMs) into three categories: non-atomic memories (NAMs), linear atomic memories (LAMs), and non-linear atomic memories (NLAMs).

The most straightforward form of NAM can be achieved by delaying light in fiber, free-space or optical resonators \cite{lvovsky2009optical, kaneda2019high, pang2020hybrid}. Active switching of transmission in these systems can enable controllable retrieval.  Other types of NAM can be achieved using optomechanical systems \cite{fiore2011storing, wallucks2020quantum} where optical coherence is mapped onto coherent oscillation of a micro or nanomechanical oscillator.

\onecolumngrid
\twocolumngrid

An ensemble of atoms (or atom-like qubits) in free space \cite{hosseini2011high, wang2019efficient, lvovsky2009optical}, inside low-Q optical resonators \cite{bao2012efficient, sabooni2013efficient, ma2022high, zhong2017nanophotonic, craiciu2019nanophotonic, craiciu2021multifunctional}, or near waveguide \cite{askarani2019storage, liu2022demand} can serve as LAM media. Coherent mapping of quantum information between flying qubits (photons) and stationary qubits (e.g. atoms, ions, superconducting qubits) can occur in the form of exchanging a single collective excitation. In such a case, a collection of atoms shares a single excitation that can be mapped to/from a flying photon. The LAM media operate in the linear regime where different optical or atomic qubits act independently of one another. In this regime, memory properties including the probability of photon emission and the amount of light’s phase shift are mostly independent of the number of atoms in the memory. A pair of entangled photons stored inside two spatially separate memories can result in the entanglement of the two atomic memories. A typical entangled photon pair generated via spontaneous process, e.g. spontaneous parametric down-conversion (SPDC) \cite{couteau2018spontaneous} or spontaneous Raman scattering (SRS) \cite{chou2004single}, can also be stored to create entanglement between quantum memories, albeit non-deterministically. There are various LAM protocols that enable the storage of flying qubits including electromagnetically induced transparency \cite{fleischhauer2000dark, phillips2001storage}, controlled reversible inhomogeneous broadening (CRIB) \cite{alexander2006photon, sangouard2007analysis}, Gradient Echo Memory \cite{hetet2008electro}, Raman storage \cite{saunders2016cavity, michelberger2015interfacing, guo2019high}, atomic frequency comb \cite{afzelius2009multimode, afzelius2010demonstration}, to name a few. As we shall see the SRS process can also be used for time-delayed photon pair generation where during the delay time, information (half of the entangled state) is stored in atoms that play the roles of both the source and the memory. This enables the implementation of entanglement between memories \cite{duan2001long}.

On the other hand, nonlinear memories or NLAMs rely on nonlinear atom-light interactions to create entanglement between photons and atoms ((see Fig. \ref{fig:fig2_QMconcept}(c))). A single atom inside a high-cooperativity optical cavity can nonlinearly interact with a photon such that an electromagnetic field emitted by the atom can interfere with itself modifying the emission rate of the atom inside the cavity. In this scenario, the spontaneous emission into the free space can be suppressed, and highly directional emission and efficient interaction can be achieved. As the result, NLAM media can enable deterministic entanglement between photons and atoms.  Instead of using the strong atom-photon interactions, NLAM can also be achieved without high-Q optical cavities and using strong atom-atom interaction realized, for example, between two Rydberg atoms \cite{li2013entanglement, li2016quantum}. In this case, spontaneous emission of high-order photon numbers, which produce noise, are suppressed via the Rydberg blockage effect, making the process of atom-photon entanglement nearly deterministic. 

\onecolumngrid
\twocolumngrid

\subsection{Absorptive LAM Protocols} 

\begin{figure}[!ht]
\centering
	\includegraphics[width=4cm]{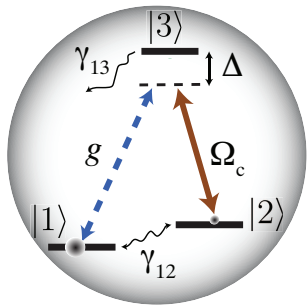}
	\caption{
	Energy-level diagram for a $\Lambda$-type atom driven by control and probe fields. For simplicity, the detuning is set to be the same for both transitions (i.e. zero two-photon detuning). $\Delta$ is one-photon detuning, $\gamma_{ij}$ is damping rate between levels $i$ and $j$.}
	\label{fig:level-diagram}
\end{figure}

In general, LAMs can be realized through an absorption or emission process. In this section, we review absorptive LAM protocols. Consider a three-level $\Lambda$-type atomic system, consisting of one ground state $\ket{1}$, a meta-stable state $\ket{2}$, and an excited state $\ket{3}$. A detuned control field of Rabi frequency $\Omega_c$ is applied on the $\ket{2}-\ket{3}$ transition while a quantum field is coupled to the $\ket{1}-\ket{3}$ transition with a coupling strength $g$, and a detuning $\Delta$, as shown in Fig. \ref{fig:level-diagram} . The total Hamiltonian of the system is $\hat{H}=\hat{H}_0+\hat{H}_I$, where $\hat{H}_0$ is the Hamiltonian of the free atom and the quantum field given by
\begin{equation}
   \hat{H}_0/\hbar=\sum_{m=1}^{N} \left(\hat{\sigma}_{33}^{(m)} \omega_{13}+\hat{\sigma}_{22}^{(m)}\omega_{23}\right) + \int \hat{a}^\dagger_k(\omega_k) \hat{a}_k(\omega_k)dk,
\end{equation}
where the sum is over $N$ atoms with atomic operators for atom $m$ described by $\hat{\sigma}_{ij}^{(m)}=\ket{i}_m\!\bra{j}$. Here $\hat{a}_k$ is the annihilation operator of the quantized field mode $k$, and $\omega_{ij}$ is the $\ket{i}-\ket{j}$ transition frequency. 
The term $\hat{H}_I$ also describes the coupling between the fields and atoms. Under the rotating wave approximation (where rapidly oscillating terms are ignored since they eventually average out to zero), and dipole approximation (where the transition matrix elements between states are written in terms of the transition dipole operator $\hat{d}$), this Hamiltonian can be written as \cite{gorshkov2007photon}
\begin{equation}
\begin{aligned}
   \hat{H}_I/\hbar= -\sum_{m=1}^{N}(\int g \hat{a}_k e^{i\omega_k t}\hat{\sigma}_{31}^{(m)} e^{i k z} dk \\ +\Omega _{c}\hat{\sigma}_{32}^{(m)}e^{-i\omega_c t+ i \phi_c }+H.c.).
   \end{aligned}
\end{equation}
Here, $g\simeq\bra{3}\hat{d}.\hat{E}_p\ket{1}/\hbar$, $\Omega_{c}\simeq \bra{3}\hat{d} . E_c\ket{2}/\hbar$, $E_c$ ($\hat{E}_p$) is the control (quantum probe) field vector, $c$ is the speed of light, $\omega_c$ ($\omega_p=\omega_k$) is the control (quantum probe) laser frequency, and $\phi_c$ is the relative driving phase of this laser.


 Assuming all atoms are equally coupled to the field, we can simplify the integration and describe a monochromatic quantum field with a slowly varying operator $\hat{\mathcal{E}}(z,t)=\int{dk\hat{a}_{k}e^{ik z}}$ that interacts with the ensemble with the effective coupling strength $\sqrt{N}g$. The Maxwell-Bloch equations in a rotating frame are then given by \cite{gorshkov2007photon, gorshkov2007photoncavity}
\begin{equation}
\begin{aligned}
\partial_t\hat{P}&=-(\gamma_{13}+i \Delta) \hat{P}+i \Omega_c \hat{S}+i g \sqrt{N}\hat{\mathcal{E}},\\
\partial_t\hat{S}&=-i\gamma_{12}\hat{S}+ \Omega^*_c \hat{P},\\
(\partial_t +&c\partial_z)\hat{\mathcal{E}}= i g\sqrt{N}\hat{P}.
\end{aligned}
\label{EOM}
\end{equation}
Here we defined $\hat{P}(z,t)=\sqrt{N}\hat{\sigma}_{13}(z,t)$ and $\hat{S}(z,t)=\sqrt{N}\hat{\sigma}_{12}(z,t)$ as polarization
and spin-wave operators, respectively, where $\hat{\sigma}_{13}(z,t)$ and $\hat{\sigma}_{12}(z,t)$ are the slowly varying collective atomic operators,  $\gamma_{13}$ ($\gamma_{12}$) is the damping or decoherence rate of the $\ket{3}\leftrightarrow\ket{1}$ ($\ket{2}\leftrightarrow\ket{1}$) transition, and $\Delta=\omega_{13}-\omega_p$ is one-photon detuning as shown in Fig.\ref{fig:level-diagram}. 

In general, it is beneficial to place the atomic system inside a cavity to realize efficient light-matter interactions. When a cavity has low loss and a small interaction volume (i.e. in the high cooperativity regime) the spontaneous emission and coupling rates can also be modified. Absorptive memories with near unity efficiency can be implemented when low-Q cavities are employed \cite{afzelius2010impedance, moiseev2010efficient, gorshkov2007photoncavity}. In the above system, if we assume the $\ket{1}-\ket{3}$ transition of atoms is coupled to a quantized cavity radiation mode, the Heisenberg equation of motion for the slowly varying cavity mode annihilation operator can be written as
\begin{equation}
\hat{\mathcal{E}}= i g\sqrt{N}\hat{P}-\kappa \hat{\mathcal{E}}+\sqrt{2\kappa}\hat{\mathcal{E}}_{in},
\end{equation}
where $2\kappa$ is the cavity decay rate, and $\hat{\mathcal{E}}_{in}$ is the cavity input field.

In the following, we review some of the widely used absorptive memory protocols used to realize LAM.

\subsubsection{Electromagnetically Induced Transparency (EIT)}
EIT memory uses a relatively intense control laser to reversibly imprint a state of light (i.e., probe field) onto an atomic state \cite{fleischhauer2005electromagnetically}. When the fields are in a two-photon resonance condition, a transparency window in the absorption spectrum of an otherwise opaque atomic system will be induced at the frequency of the probe laser.  {As soon as the probe is spatially compressed inside the medium (since its group velocity is greatly reduced),} its coherence can be adiabatically transferred from the atomic polarization mode to the spin mode by progressively reducing the control field power to zero (i.e., bringing the probe field to a complete halt) \cite{liu2001observation}. To retrieve the stored photon, one needs to gradually turn on the control field. The group velocity of light inside the medium is proportional to the control laser power and inversely proportional to the optical density, $d$. In contrast, the EIT bandwidth or width of the transparency window is (inversely) proportional to (control power) $\sqrt{d}$. Thus the delay-bandwidth produces scales with $\sqrt{d}$. At high optical densities, however, one needs to be careful to reduce the four-wave mixing (FWM) noise \cite{lauk2013fidelity}, for example by using cavities or laser-cooled atoms.

\subsubsection{Autler-Townes Splitting (ATS)}
Similar to the EIT, the ATS is an optically controlled memory \cite{saglamyurek2018coherent}. However, the mapping process of the latter has a non-adiabatic (i.e., fast) nature as we turn off the control field abruptly. Here the probe field is absorbed through the ATS peaks \cite{autler1955stark}, generated by the ac-Stark splitting, and its coherence is mapped into the spin mode through a non-adiabatic exchange process mediated by the polarization mode. 
As a non-adiabatic memory protocol, ATS is more flexible in terms of the required control field power and optical depth compared to the adiabatic protocols \cite{rastogi2019discerning}. In principle, the smaller the optical depth and control field power, the less four-wave mixing noise. However, in turn, ATS is only efficient for the storage of input fields whose bandwidths (at full-width half-maximum) are larger than the linewidth of the transition in resonance with the probe field (i.e., referred to as broadband regime).

\subsubsection{Off-resonant Raman}
The detuned version of the EIT memory when both control and probe fields are detuned from the excited level, as shown in Fig.\ref{fig:level-diagram}, is called off-resonant Raman memory \cite{reim2010towards, reim2011single, saunders2016cavity}. 
This scheme is also established based on the adiabatic elimination of the polarization mode. However, here, the Raman scattering causes the storage of the probe field. Compared to the EIT, the detuned nature of the Raman makes it less susceptible to fluorescence noise and enables large-bandwidth storage possible. These benefits, however, come at the expense of a stronger control field to compensate for the weak coupling caused by the off-resonant operation \cite{saglamyurek2018coherent, nunn2007mapping, gorshkov2007photon}. As a result, the four-wave mixing noise is typically higher in Raman memory than in EIT quantum memory. An effective way to suppress the FWM noise is to use a cavity in resonance with the probe field but antiresonance with the noise field \cite{nunn2017theory}. 

\subsubsection{Controlled Reversible Inhomogeneous Broadening (CRIB)}
CRIB is an engineered absorption-based quantum memory \cite{alexander2006photon}. This protocol requires the generation of a narrow spectrally isolated absorption line (i.e., spectral tailoring) from an inhomogeneously broadened line. This can be done via the optical pumping of the rest of the population into an auxiliary energy level. An external magnetic or electric field gradient is then used to broaden the absorption line controllably and reversibly. For example, broadening can be accomplished with an electric field gradient and the linear Stark effect in rare-earth doped solids with a permanent dipole moment \cite{lauritzen2010telecommunication}. This results in the absorption of the probe field by the broadened line. Following the absorption, the external field is turned off, allowing the population to be transferred to a ground state using an optical pulse. Re-emission of the stored photon can occur later by transferring the population back to the excited state and reversing the sign of the external field to create a photon echo.
When the propagation direction of the probe is parallel to the direction of the external field variation, the resulting scheme is called gradient echo memory (GEM) \cite{hetet2008electro} or longitudinal variant of CRIB that enables on-demand retrieval. The GEM protocol can be very efficient \cite{hedges2010efficient,hosseini2011high} as it suppresses re-absorption, and it allows for arbitrary and coherent control of optical information \cite{hosseini2009nature, sparkes2012precision, campbell2014configurable}. 

In contrast to EIT or Raman memories whose multi-mode capacity scales as $\sqrt{d}$, CRIB is a memory protocol with a multi-mode capacity that scales linearly with the optical depth \cite{nunn2008multimode}. 

\subsubsection{Atomic Frequency Comb (AFC)}
AFC is another engineered absorption-based scheme that has attracted considerable attention \cite{afzelius2009multimode}. Unlike the CRIB protocol, which only needs a single narrow absorption line, AFC requires many periodic absorption lines (a comb), in which each pair of nearest-neighbor peaks is separated by a frequency $\delta$. Such a comb can be prepared by optical pumping techniques to selectively transfer atoms into a meta-stable state (i.e., tailoring the absorption profile). Here, the bandwidth of the photon to be stored is limited from below (above) by peak spacing (comb overall width). In this memory protocol, after a specific time determined by $1/\delta$, the rephasing of atomic excitations results in the automatic re-emission of the photon. To address the limitation of pre-determined re-emission, optical control fields can be employed to temporarily map the coherence of the stored optical excitation to a long-lived ground state as a spin-excitation mode. This also results in longer storage times. Alternatively, combining CRIB and AFC protocols for each absorption line can also make the retrieval on-demand \cite{lauritzen2011approaches}. Another approach is to use DC Stark shift to create a $\pi$ phase difference in the atomic ensemble, later apply another electric field to rephase the system to retrieve the photon \cite{craiciu2021multifunctional, horvath2021noise, liu2020demand}.

It has been shown that, unlike many other memory protocols, the multi-mode capacity of the AFC is not limited by the optical depth \cite{nunn2008multimode}. To be more precise, the ratio of the excited state storage time to the duration of the stored pulse determines how many temporal modes can be stored in an AFC quantum memory. This ratio is independent of the optical depth, making AFC a promising candidate for designing multi-mode memories. 


\subsubsection{Other Protocols}
Besides the commonly used protocols listed above, there are more memory schemes. The traditional two-pulse photon echo techniques suffer from spontaneous emission and gain due to population inversion onto excited states. 4-level photon echo schemes \cite{beavan2011photon} are proposed to mitigate the issues by using more energy levels and spectral filtering, but imperfect $\pi$ pulses still leave some atoms in the excited states, which makes them difficult to implement at the single-photon level. The Revival of Silenced Echo (ROSE) is another technique\cite{damon2011revival} in which spatial phase mismatch is created between an input pulse and first rephasing $\pi$ pulse to suppress the first echo. Sending another $\pi$ pulse then rephases the atomic ensemble. This method also suffers from imperfect $\pi$ pulses. The Noiseless Photo Echo technique\cite{ma2021elimination} has been proposed by combining 4-level photon echo and ROSE. Stark Echo Modulation Memory (SEMM) \cite{arcangeli2016stark, fossati2020frequency} has also been introduced by utilizing inversion symmetry of linear Stark effect in solids, which can create a $\pi$ phase difference among two types of frequency shifts in the atomic ensemble, instead of using spatial phase mismatching. Four-wave-mixing in the atomic ensemble can also be used for delaying or storing quantum light \cite{marino2009tunable}. Two-photon off-resonant cascaded absorption (ORCA) has been used to demonstrate intrinsically noise-free quantum memory \cite{kaczmarek2018high, thomas2022single}. Using off-resonant Faraday interactions in warm vapor cells, ms-long quantum storage has been demonstrated \cite{julsgaard2001experimental, julsgaard2004experimental} 
Recently, fast ladder memory (FLAME) was demonstrated to store light into electronic orbitals of rubidium vapor \cite{finkelstein2018fast}. It was also shown that coherence can be mapped to noble-gas spin via spin-exchange collisions to reach hour-long coherence in hot atomic vapor \cite{katz2021coupling, katz2022optical, shaham2022strong}. 

\subsection{Emissive LAM protocols}

In emissive memories, the excitation is heralded by the emission of an idler photon controlled by an external laser pulse. The spontaneous emission of the idler photon detected in a certain detection mode gives rise to the creation of stored atomic excitation that is entangled with the idler photon. Subsequent retrieval of the atomic excitation can be obtained by reversing the process using a strong laser thus creating a signal photon entangled with the idler.  

When a spontaneously scattered photon is emitted into the detection mode, the ensemble $N$ of atoms initially in the ground state,  $| G \rangle=| g_1g_2 ...g_{N}\rangle $,  is transferred to a collective excitation $| S \rangle= \frac{1}{\sqrt{N}}\sum_i{| S_i \rangle}$, where $| S_i \rangle=| g_1g_2..s_i..g_N \rangle$ is the atomic state when atom $i$ is excited to the meta-stable state $| s_i \rangle$. The summation over $i$ denotes the sum of all permutations. The state of the system can then be written as an entangled state $| \psi \rangle=a| G,0 \rangle+b| S,1 \rangle$, where $0$ and $1$ denote zero and one photon states and $a$ and $b$ are the normalization coefficients. A similar process can be used to generate polarization photonic qubits as shown in Fig.\ref{fig:Fig4_DLCZ_QED}(a). 

Using the above-mentioned process, the DLCZ scheme \cite{duan2001long} was first proposed as a means for long-distance quantum communication. In this scheme, an optical laser field is used to establish a Raman transition between the two ground states of a $\Lambda$-type atomic ensemble. This results in a collective excitation mode within the ensemble. Later, a second laser can be used to retrieve the stored collective excitation as a propagating field. Idler photons from two distant ensembles can then be directed to a 50:50 beam splitter (BS) to perform join measurement. The latter erases the which-path information of photons. The detection of a single-photon after BS results in the generation of entanglement between the ensembles (see Fig.\ref{fig:Fig4_DLCZ_QED}(a) ). The recalled photons from two neighboring ensembles can then be directed to another BS. A successful Bell state measurement results in entanglement swapping between the two pairs of entangled ensembles, by which entanglements can be extended to long distances.

In 2003, several groups proposed a two-photon interference scheme that can also be used to generate entanglement between remote qubits \cite{simon2003robust,duan2003efficient, feng2003entangling}. In this scheme, qubits are initially prepared in the excited state, which can later decay into either of the ground states. As a result, an entangled state between the emitted photon and the qubit can be created. Two emitted photons, each coming from a qubit, meet at a BS located in between qubits for a Bell state measurement. The detection of two photons by two different detectors projects the joint state of distant qubits into a Bell state. 

A single-photon interference scheme can also be implemented to achieve an ``event-ready" Bell test \cite{simon2003robust}. Moreover, a double-heralded single-photon interference scheme can also generate entanglement between distant qubits \cite{barrett2005efficient}. 
 Using this scheme, the first loophole-free Bell test has been demonstrated \cite{hensen2015loophole}. 

\subsection{Nonlinear Atomic Memories (NLAM)}
The LAM protocols discussed so far rely on absorptive or emissive processes integrated with spontaneous photon generation to create probabilistic photon-atom or atom-atom entanglement.   In this section, we discuss the principles of NLAM in the context of strong light-atom interactions inside high-Q cavities and strong atom-atom interactions based on Rydberg atoms, as methods for near-deterministic generation of entanglement.
\subsubsection{Strong cavity interaction}
Photon-atom interactions can be enhanced by optical resonators to overcome the probabilistic limitation in LAMs. Such interaction is key to deterministic entanglement creation between atomic and photonic qubits, as well as superconducting qubits, and are used in various quantum technologies. We note that simply placing an atom inside an optical cavity does not make its interaction with a photon less probabilistic. Certain interaction parameters need to be engineered to reach the strong coupling regime. We refer to this interaction regime as high cooperativity regime , where an atom or a photon state can interfere with itself. If losses are small and interactions can happen faster than the decay rate of the atom, the EM field of the photon can interfere with the atom and suppress or enhance the emission. The interaction is therefore nonlinear at the single particle level. 

The interaction parameter known as cooperativity, $C$, is defined as the ratio of the square of the light-atom interaction rate to the loss rates by atom and cavity $C=g^2/\gamma\kappa$. On the atomic and cavity resonance, the ratio of the scattered power, $P_{4 \pi}$, and the transmitted power, $P_{tr}$ to the incident power, $P_{in}$, can be expressed in terms of the cooperativity \cite{tanji2011interaction} as $P_{4\pi}/P_{in}=2C/(1+C)^2$, and $ P_{tr}/P_{in}=1/(1+C)^2.$

\begin{center}
\begin{figure}[!ht]
	\centerline{\includegraphics[width=1\columnwidth, angle=0]{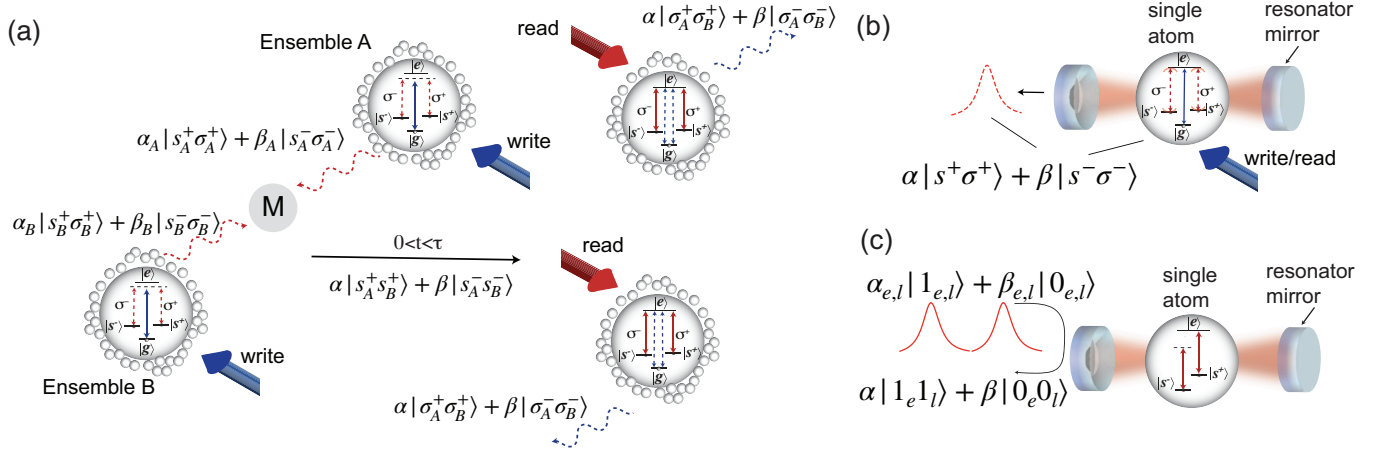}}
	\caption{
	(a) An ensemble of atoms can generate polarization qubits via spontaneous Raman scattering process while entangled with a single collective spin excitation stored in the ensemble.  Joint measurement (M) of two photonic qubits (red) generated from distant ensembles via two “write” laser fields can project the two ensembles into an entangled state. Collective excitations of the two ensembles can be mapped to flying entangled qubits (blue) after a desired storage time. (b) The probability of atom-photon entanglement in (a) can be enhanced using high-cooperative interaction of a single atom inside an optical resonator. (c) Another scheme for photon-atom and photon-photon entanglement is shown where the reflection of a photonic qubit off a high-cooperative cavity containing a single atom can create near-deterministic atom-photon entanglement. Subsequent reflection of a second photonic qubit and detection of atomic state can create near-deterministic photon-photon entanglement.}
	\label{fig:Fig4_DLCZ_QED}
\end{figure}
\end{center}

It can be seen that at high cooperativities, the transmission and scattering are suppressed to the point where the photon is reflected from the cavity with high probability. This however does not indicate that the photon will not enter the cavity or not interact with the atom. It is the interaction with the atom that gives rise to such ``cavity blocking". Conversely, when the intra-cavity atom is excited, it can create an intra-cavity photon with a near-unit probability scaling with $C^2/(1+C)^2$. The strong interaction in this form can be used to create near-deterministic atom-photon or photon-photon entanglement. As seen in the example in Fig. \ref{fig:Fig4_DLCZ_QED} (b), directional Raman scattering of an optical polarization qubit  $\alpha|\sigma^+\rangle+\beta|\sigma^-\rangle$ can be entangled with the atomic spin states, $|s^+\rangle$ and $|s^-\rangle$. In the reflection scenario (example shown in \ref{fig:Fig4_DLCZ_QED} (c)), an incident photonic time-bin qubit is only reflected when an intra-cavity atom is in a certain ground state with maximum interaction. As the result, after reflection of the first qubit, the atom and photon are in an entangled state $| \psi \rangle=| 1_e,s^+ \rangle+| 1_l,s^- \rangle$, where $l$ and $e$ refer to the early and late time slots of the photonic qubit. Reflection of the second photon followed by detection of the atomic state can map the state of the two photonic qubits into an entangled state. Such entanglement can be ideally deterministic in the low-loss and high-cooperativity regimes.

\subsubsection{Strong atom-atom Rydberg interaction}


Strong interaction between atoms excited to Rydberg states can also enable near-deterministic light-atom entanglement. The effect of Rydberg-level interactions in various quantum information applications is usually explained in terms of the Rydberg blockade mechanism, where the presence of an excited Rydberg atom blocks nearby atoms from being excited. However, such a picture of the blockade mechanism is not strictly correct to describe ensemble-light entanglement. This can be seen by considering two atoms in a molecular basis. For an incident field that is resonant with the Rydberg transition in an isolated atom, the single-atom excitation state is resonant, whereas the doubly excited state is shifted out of resonance. Thus, a single atom is never excited in this scheme, only a single excitation is shared by the two atoms. For $N$ atoms, the molecular picture remains valid, even if the level scheme becomes complicated. Ideally, the excitation dipole blockade can produce a collective single excitation - a Dicke state - with unit probability. As in any ensemble-based quantum memory, such a collective state is imprinted with a phase pattern that allows for its efficient mapping into a propagating single-photon wavepacket.

It is worth noting that Bariani {\it et al.} have proposed a spin-wave dephasing mechanism that achieves much the same goals \cite{bariani2012dephasing}. This approach has been used to describe the generation of quantum light and atom-light entanglement in various experiments \cite{dudin2012strongly,li2013entanglement,li2016quantum}. In this approach, a short excitation pulse creates a multiple-excitation state. Following the excitation all state components but the singly-excited one undergo dephasing as a result of the distribution of the interaction shifts $\Delta_{jj^\prime}$'s between pairs of atoms $j$ and $j^\prime$. If the $\Delta_{jj^\prime}$'s were all equal there would be no dephasing and the associated loss of the retrieved signal. In effect, the dephasing mechanism exploits interaction-induced phase factors to isolate the singly-excited component in the directional (phase-matched) optical retrieval process.
    
If interactions can be neglected in the excitation process, the atoms are prepared in a factorized state, for which the maximum population of a single collective excitation state produced via the dephasing mechanism in the storage period is limited to $1/e$ \cite{bariani2012dephasing}. In contrast, the Rydberg excitation blockade, in principle, allows one to reach unity efficiency of the collective single excitation \cite{ornelas2020demand}. 

In experiments where the prepared atomic state is intended to be mapped into a light field, the efficiency of the mapping is just as important as the atomic state preparation efficiency. The mapping efficiency is a function of cooperativity parameter $C$ (for cavity settings) or, for free-space settings, optical depth $d$ which for an atomic sample of length $L$ scales as $\sim \rho \lambda ^2 L$. 
To achieve near-unity atom-light mapping, the condition $d \gg 1$ must be achieved, which implies $L \gg (\rho \lambda ^2)^{-1}$.
The atomic density $\rho$ in its turn must be kept sufficiently low (in practice $\leq 10^{12}$ cm$^{-3}$) so as the rate of ground-Rydberg decoherence is not prohibitive. Taken together, these considerations set limits on the size and the density of the atomic sample. Thus, regardless of the values of $g_{atoms}^{(2)}(0)$ produced by the excitation blockade, interaction-induced dephasing in both the excitation
and storage phases can lead to a value of $g^{(2)}(0)/g_{atoms}^{(2)}(0)\ll 1$. As a consequence interaction-induced dephasing is an important mechanism
for the reduction of $g^{(2)}(0)$ and for entanglement generation.


\onecolumngrid
\twocolumngrid

\subsection{Multiplexed quantum memories}
\begin{center}
\begin{figure}[!ht]
	\centerline{\includegraphics[width=1\columnwidth, angle=0]{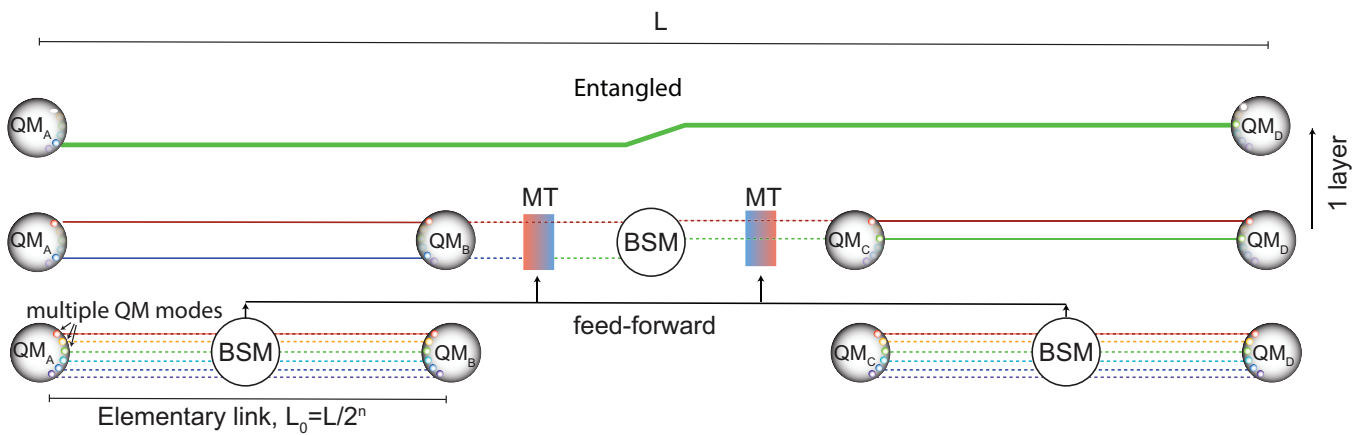}}
	\caption{
	An example of a linear network consists of quantum memories (QM) and Bell-state measurement (BSM) stations. Entanglement creations at elementary links and one layer of entanglement swappings (nested layer n=1) are used to create end-to-end entanglement between memories separated by distance L. Quantum memories are considered to support multiple atom-photon entanglement   modes (temporal, spectral, or spatial) shown by different colors. Results of BSM at the elementary links can be fed-forward to the next layer to perform mode transformation (MT) on remaining entangled photons followed by BSM on corresponding modes.}
	\label{fig:fig5_layers_MX}
\end{figure}
\end{center}

 Consider a small linear system of memories and sources as shown in Fig. \ref{fig:fig5_layers_MX} where photon-memory entanglement can occur with probability $p$. The success probability of entanglement creation between two adjacent memories separated by distance $L_0$ after Bell-state measurement (BSM) is given by $P_0 = p\eta_t\eta_d$, where $\eta_t$ and $\eta_d$ are the efficiencies of channel transmission and detection, respectively. The transmission efficiency of the fiber link is $\eta_t=e^{-L_0/2L_{att}}$, where $L_{att}=22km$ (assuming 0.2 dB/km fiber loss). Detection efficiency is bounded by BSM probability, i.e. $\eta_d$. For spontaneous sources, $p\backsim0.05$ is limited by the probability of having higher-order photon numbers. $\eta_m$ is the memory efficiency. In a multilayer network and for $n$ levels of entanglement swapping (nesting layers) in a repeater network, the entanglement rate \cite{sangouard2011quantum} can be expressed as 
 
\begin{equation}
    R_e = \frac{c}{L_0}\frac{\eta_t\eta_d^{n+3}\eta_m^{n+2}p}{3^{n+1} \prod_{k=1}^n [2^k-(2^k-1)\eta_m\eta_d)]}\
\end{equation}

As an example, for 8 memories (n=2, 4 elementary links) extending the communication distance to $L = 2^2L_0 =1000km$, assuming both memory and detection efficiencies to be 0.9, one entanglement event can be achieved around every 1 hour! Although this entanglement rate is much faster compared with direct fiber links (less than one event every $10^{10}$ seconds), in this traditional approach, due to the low entanglement rate and need for quantum memories with hour coherence times, such a network has limited practical use.

\onecolumngrid
\twocolumngrid

To go beyond such traditional communication schemes, provided the technology is available, one can simultaneously process multiple modes of photons using multimode BSMs, multimode sources, and memories \cite{sangouard2011quantum, simon2007quantum}. The photon generation probability can, in principle, approach unity where strong light-atom interactions \cite{li2013entanglement} are employed. Considering strong cavity interactions, $p$ can increase by more than an order of magnitude. Moreover, assuming $m$ (spatial, temporal or spectral) modes processed in parallel, the communication rate can linearly increase with $m$. 

Instead of parallel processing of multiple modes $m$, multiplexed processing \cite{collins2007multiplexed} of entanglement can be implemented to enhance the communication rate. Multiplexed quantum optical processing refers to the ability to perform Bell state measurement (BSM) on the multiple numbers of modes shared between two nodes. The results are then fed forward to perform subsequent measurements at the next layer, only between the entangled modes established at the first layer. Therefore, multiplexing relies on switching under feed-forward control, a crucial functionality that benefits from the co-existence of classical signals in the channel. In the limit of low memory lifetimes, the communication rate scales as $(mP_0)^{\text{2$^{\text{n}}$}}$ for $mP_0<\!\!<1$, while it is $mP_0^{\text{2$^{\text{n}}$}}$ for parallel processing ($n$ is the number of nested layers in the network). The rate scaling is also significant in the intermediate memory lifetimes (a few 100 ms) resulting in practically useful communication rates. According to Ref. \cite{collins2007multiplexed}, in the high-memory-lifetime limit, the multiplexing only marginally improves the rates when compared to parallel schemes.

\subsection{Quantum memory properties}


The performance of quantum memories can be assessed by different figures of merit discussed below.
The fidelity of memory shows how close the retrieved quantum state is to the input quantum state by quantifying the overlap between them. In general, infidelity can happen due to different mechanisms such as loss, four-wave mixing noise, decoherence through interactions, and thermal effects. 

In some memory protocols where we aim for single-photon storage, the completion of the scheme is conditioned on the retrieval of a specific outcome (e.g., the re-emission of a single photon). One can then define a conditional fidelity, which quantifies the overlap of the emitted single-photon state with the input single-photon state.  

Memory efficiency quantifies the energy of the output compared to the input, or more specifically, the success probability of recovering a photon. The efficiency can be reduced by different types of loss present in any system (e.g., absorption, scattering). In ensemble-based systems, the collective interference effect boosts efficiency. On the other hand, in single atomic systems, using cavities can enhance light-matter interaction and therefore efficiency. For certain protocols, the re-absorption of the re-emitted light can limit the maximum efficiency to below $54\%$ \cite{afzelius2009multimode}. In backward retrieval or using impedance-matched cavities, however, efficiency can be near-unity. 

 { In addition to communication times as an important factor influencing achievable repetition rates of repeater protocols, the bandwidth of a quantum memory also }determines the achievable communication rate through the multi-mode capability and repetition rates. In principle, the longer the duration of a pulse, the smaller the achievable bandwidth. The maximum attainable bandwidth depends on the memory protocol and characteristics of the storage system. 

Storage time is another important key factor of memory that can be limited by medium's coherence time. In particular, in long-distance quantum communication applications, the memory storage time should be at least comparable with the entanglement generation time  {e.g., the time required to establish entanglement between two distant nodes in addition to the time it takes for a classical signal to go back over the same distance and notify us about the outcome of measurements.} To increase the storage time in atomic systems, one can use zero-first-order Zeeman transitions that have zero gradient with respect to the magnetic field and are therefore less sensitive to the field fluctuations \cite{zhong2015optically}. Additionally, in most atomic systems, lowering the temperature can slow down decoherence processes and therefore, extend the storage time. Dynamical decoupling techniques such as Carr-Purcell-Meiboom-Gill (CPMG) sequences \cite{carr1954effects, meiboom1958modified}, can also suppress noise effects in a given system \cite{siyushev2014coherent, souza2011robust}. 

The selection of the wavelength is mainly influenced by the memory application and memory medium. For example, telecom wavelengths are recommended for communication through optical fibers as they have minimal absorption during transmission \cite{asadi2020protocols}. On the other hand, for free-space communications, diffraction is the main source of transmission loss \cite{boone2015entanglement}. 

In general, the wavelength of the recalled photon can be changed using nonlinear interactions or transduction mechanisms \cite{de2012quantum, lauk2020perspectives, asadi2022proposal,chen2020efficient}. By using difference-frequency or sum-frequency generation processes, the wavelength can be converted to the telecom wavelength for long-distance propagation in fibers \cite{albrecht2014waveguide, allgaier2017highly, morrison2021bright, dreau2018quantum}. It is also possible to utilize cold atoms for wavelength conversion \cite{walker2018long, bock2018high}. Telecom quantum memory with hot Rb vapor has also been demonstrated by cascaded two-photon absorption \cite{thomas2022single}. Furthermore, to avoid limitations in memory platform selection, photon-pair sources can be used to separate entanglement generation and storage processes \cite{sangouard2011quantum, rakonjac2021entanglement, lago2021telecom}. 

Multi-mode capacity is the capacity to store independent quantum states simultaneously in a single memory. Multiplexing is especially important for improving communication rates over long distances \cite{sangouard2011quantum}. Multiple degrees of freedom can be employed in a multi-mode memory, including, temporal, spectral, spatial, and angular. To further expand the mode's capacity, some of these degrees of freedom can also be combined \cite{yang2018multiplexed}.


\section{State of the art}

The efficient and deterministic distribution of quantum entanglement is key to developing future quantum networks \cite{wehner2018quantum}. To date, several entanglement-distribution demonstrations have been carried out with and without memories, namely: $3-$node quantum entanglement distribution using cold atomic memories with overall efficiency $10^{-8}$ \cite{jing2019entanglement}, memory-less ground-space integrated quantum key distribution (QKD) network over 4600km \cite{chen2021integrated}, memory-based QKD with rate enhancement relying on asynchronous BSMs \cite{bhaskar2020experimental}, 8-node all-optical quantum communication with probabilistic bi-party entanglement distribution \cite{wengerowsky2018entanglement, joshi2020trusted}, and entanglement of solid-state QOMs \cite{liu2021heralded, lago2021telecom, pompili2021realization}. Faithful quantum state transfer between a cold atomic ensemble and a rare-earth-doped crystal has been demonstrated \cite{maring2017photonic}. Built-in nuclear spin quantum memory in nitrogen-vacancy (NV) centers in diamond has been used for quantum teleportation through a middle node \cite{hermans2022qubit}.

Below we discuss the most advanced QOMs and their application in the context of quantum communication networks. Table \ref{fig:table_QM} summarized the most recent significant advances in the development of stand-alone QOMs in different platforms. We also summarize recent entanglement distribution demonstrations using LAMs and NLAMs between separate nodes in Table \ref{fig:table_QN}.


\subsection{Advances in absorptive quantum memories}
Many platforms are being used to develop absorptive LAMs, including rare-earth ions doped crystals \cite{ma2021one, saglamyurek2011broadband, businger2022non, liu2021heralded, lago2021telecom, hedges2010efficient}, cold atomic systems \cite{saglamyurek2018coherent, hsiao2018highly, wang2019efficient, dudin2013light, chang2019long}, and hot vapors \cite{novikova2007optimal, novikova2008optimal, chrapkiewicz2017high, phillips2008optimal, katz2018light}. Below, we summarize the most notable results.

Following the first demonstration of an EIT memory in a cold atomic cloud \cite{liu2001observation}, several experiments were carried out to improve light storage using this protocol. In particular, a storage efficiency of $92\%$ \cite{hsiao2018highly}, and a conditional fidelity beyond $99\%$ (with an efficiency of $85\%$) \cite{wang2019efficient} have been achieved using cold cesium (Cs) and rubidium (Rb) atoms, respectively. 

Using off-resonant Raman memory, an efficiency of up to $82\%$ with an unconditional fidelity of $98\%$ and a bandwidth in the range of $10^2$ MHz, has also been reported in Rb vapor \cite{guo2019high}. 

Four-wave mixing noise is an important source of infidelity in Raman memories (especially those operating at room temperature) that can be suppressed by engineering interference \cite{thomas2019raman} or using cavities to suppress the noise \cite{nunn2017theory,saunders2016cavity}.

With respect to the multimode capacity, an angularly multiplexed holographic memory based on the Raman protocol has been demonstrated using 60 spin-wave modes in warm Rb vapors \cite{chrapkiewicz2017high}.

Similar to the Raman scheme, ATS offers storage of broadband input fields. In particular, optimal ATS memory in the broadband regime requires an optical depth that is nearly 6 times smaller than the one required for an optimal EIT scheme \cite{saglamyurek2021storing}. Using the ATS memory, a bandwidth of 14.7 MHz \cite{saglamyurek2018coherent}, and an efficiency of up to $30\%$ \cite{saglamyurek2021storing} have been demonstrated in cold Rb atoms and Bose-Einstein condensate (BEC) platforms, respectively. Recently, theoretical research on cavity-enhanced ATS quantum memory using T centers has also yielded promising results \cite{higginbottom2022memory}.


\onecolumngrid
\begin{center}
\begin{table}[h]
\scriptsize
\begin{tabular}{l c c c c c c}
   & \cellcolor{cyan}Diamond defects & \cellcolor{yellow}RE crystal (Er$^{3+}$)& \cellcolor{yellow}RE crystal (Pr$^{3+}$)& \cellcolor{yellow}RE crystal (Eu$^{3+}$)& \cellcolor{green}Laser-cooled atoms & \cellcolor{red}Hot vapor \\
  \hline
  \hline
  $\mathcal{F}$ & $0.89^a$ , $0.94^b$ & $0.98^c$& $0.98^e$& $>0.95^{g,h}$& $0.99^i$, $>0.95^j$ & $0.97^k$\\
  \hline
  $\eta$ & NR & $0.07^c$, $0.22^d$& $0.69^e$, $0.56^f$ & $0.1^g$& $0.68^i$,$>0.78^j$ & $0.67^k$\\
  \hline
  $\tau_s$ & 13ms$^a$, 75s$^b$  & 320ns$^c$, 660ns$^d$& 1.2ms$^e$, \SI{1.1}{\micro\nothing}s$^f$& \SI{21}{\micro\nothing}s$^g$, 1h$^h$& \SI{1.2}{\micro\nothing}s$^i$, \SI{3}{\micro\nothing}s$^j$ & 100ns$^k$\\
  \hline
  $\Delta\omega$(MHz) & NR & 100$^c$, 6$^d$ & 0.14$^e$, 11$^f$ & 700 $^g$, 1$^h$ & $\sim$1$^i$, 3.3$^j$ & $<$1$^k$\\
  \hline
  T (K) & 0.1$^a$, 3.7$^b$ & 0.1$^c$, 1.5$^d$ & 3$^e$, 2.1$^f$& 3.5$^g$, 1.7$^h$ & $\SI{20}{\micro\nothing}^i$, 0.2m$^j$ & 368$^k$\\
  \hline
   B (T) & 0.3$^a$, 0.04$^b$ & 1$^c$, 7$^d$ & 0$^{e,f}$& 0$^g$, 1.3$^h$ & 0$^{i,j}$ & 0$^k$\\
  \hline
  $\lambda$(nm) & 737$^a$, 637$^b$& 1536$^{c,d}$& 606$^{e,f}$ & 580$^{g,h}$& 795$^{i,j}$ & 795$^k$\\
  \hline
  MC & NR &1650$~^l$ & $10^5$(est.)$~^m$  & $>$100$~^n$ & 225$~^o$  & 60$~^p$\\
  \hline
  
\end{tabular}
\caption{
	Summary of most recent advances in stand-alone quantum memory developments in different platforms highlighting measured fidelity ($\mathcal{F}$), efficiency ($\eta$), storage time ($\tau_s$), bandwidth ($\Delta\omega$), operating temperature (T), applied magnetic field (B), wavelength ($\lambda$), and multimode capacity (MC). Corresponding references are a: \cite{sukachev2017silicon}, b: \cite{bradley2019ten}, c: \cite{liu2022demand}, d: \cite{stuart2021initialization}, e: \cite{hedges2010efficient}, f: \cite{sabooni2013efficient}, g: \cite{ma2021elimination}, h: \cite{ma2021one}, i: \cite{vernaz2018highly}, j: \cite{wang2019efficient}, k: \cite{ma2022high}, l:  \cite{wei2022storage} , m: \cite{yang2018multiplexed}, n: \cite{ortu2022storage}, o: \cite{chang2019long}, and p:  \cite{chrapkiewicz2017high}. Promising results were also obtained with other rare-earth ions such as Tm$^{3+}$ \cite{askarani2021long}, Nd$^{3+}$ \cite{liu2021heralded} and Yb$^{3+}$ \cite{businger2020optical} (not listed in this table). }
	\label{fig:table_QM}
\end{table}
\end{center}
\twocolumngrid


For a two-level gradient-echo memory, low noise storage of optical pulses with an efficiency of $69\%$ has also been obtained in  praseodymium (Pr$^{\text{3+}}$) doped a crystal \cite{hedges2010efficient}. However, the storage time in this experiment was only a few microseconds. Later, using a three-level $\Lambda$ structured warm rubidium vapor, the recall efficiency of the GEM protocol improved to $87\%$, but the storage time remained in the same regime \cite{hosseini2011high}. In addition, using the same platform a recall fidelity of up to $98\%$ has been demonstrated for coherent pulses with around one photon \cite{hosseini2011unconditional}. The same efficiency for storage (i.e., $87\%$) has also been reported in an ensemble of laser-cooled Rb atoms with storage times of up to $0.6$ms, albeit with a low bandwidth of 210 kHz \cite{cho2016highly}. 

Temporal multi-mode storage using the CRIB has also been studied. It has been shown that to achieve an efficiency of $90\%$ using 100 temporal modes, an optical depth of around 3000 is required \cite{simon2007quantum, nunn2008multimode}. Although it is an improvement compared to multi-mode memories based on EIT and Raman, the required value of optical depth is still unrealistic.

 {Using AFC memory}, a storage bandwidth of 5 GHz has been reported in a thulium-doped lithium niobate waveguide \cite{saglamyurek2011broadband}. In ref \cite{zhou2012realization}, nearly unit conditional fidelity ($99.9\%$) was demonstrated using an Nd$^{\text{3+}}$-doped crystal. 
Besides, using the linear Stark effect, an extension of the AFC protocol (i.e., Stark-modulated atomic frequency comb) with a recall efficiency of $38\%$ and a short storage time of  0.8 $\mu s$ has been realized \cite{horvath2021noise}. Later, using the same approach, a conditional storage fidelity of $99.3\%$ have been demonstrated \cite{liu2020demand}. However, in this experiment, the storage time was limited to 2 $\mu s$ mainly due to the electrically induced broadening of the absorption peaks. In a more recent experiment, using the dynamical decoupling technique and ZEFOZ transitions of Eu$^{\text{3+}}$-doped crystal, the same group reported a storage time of 1 hour and a fidelity of $96.4\%$ \cite{ma2021one}. Spin-wave storage has been demonstrated in Eu$^{\text{3+}}$ \cite{jobez2015coherent, ma2021one, ortu2022storage, holzapfel2020optical} and Pr$^{\text{3+}}$ \cite{gundougan2015solid, ferguson2016generation, yang2018multiplexed, rakonjac2021entanglement}, and more recently in Kramers ion ytterbium (Yb$^{\text{3+}}$) \cite{businger2020optical}. Non-Kramers ions such as Pr$^{\text{3+}}$ and Eu$^{\text{3+}}$ doped in crystals have been used to demonstrate long-time and high-efficiency quantum storage due to their long lifetime and coherence time, since their electronic spins can be quenched by crystal fields, while Kramers ions such as Er$^{\text{3+}}$ and Yb$^{\text{3+}}$ suffer from short lifetime and coherence time due to spin flips and cross-relaxation. Strong magnetic field ($>$1T) and low temperature ($\sim$1K) can greatly increase the excited state coherence time of Er$^{\text{3+}}$ doped into yttrium orthosilicate to be 1ms \cite{bottger2009effects}. At the similar conditions, hyperfine state with coherence time over 1s has been reported at Ref. \cite{ranvcic2018coherence}. Efficient state preparations have also been developed by the same group, which has achieved $22\%$ storage efficiency after 660ns \cite{stuart2021initialization}. More effective optical pumping methods need to be investigated to prepare atomic frequency combs with better structures and completely empty the hyperfine levels for spin-wave storage. 

Cavity-assisted AFC memory allows for highly efficient storage of quantum states. Several experiments using the impedance-matched cavity approach \cite{afzelius2010impedance, moiseev2010efficient} have demonstrated cavity-enhanced AFC memory with efficiencies of up to $56\%$ in rare-earth ions doped solids \cite{sabooni2013efficient, sabooni2013cavity, jobez2014cavity, akhmedzhanov2016cavity, davidson2020improved}.

The inhomogeneously broadened absorption profile of rare-earth ions doped into solids has been utilized to demonstrate storage of 100 temporal modes in Nd-doped crystal \cite{tang2015storage}. A more recent experiment used Yb$^{\text{3+}}$ doped into a crystal for multiplexed storage of 1,250 temporal modes with a bandwidth approaching 100 MHz \cite{businger2022non}. Although temporal multiplexing is well suited for the AFC memory \cite{afzelius2009multimode}, spectral and spatial multiplexing can also be performed using this protocol. The storage of a polarization qubit using two spatially separated ensembles of Pr$^{\text{3+}}$ doped into a crystal has been demonstrated\cite{gundougan2012quantum}. In addition, multiplexed storage of weak coherent states into up to 26 spectral modes has been reported in a Tm$^{\text{3+}}$ doped waveguide \cite{sinclair2014spectral}. More recently, simultaneous storage of 15 frequency bins of a photon pair, where each spectral bin contains 9 temporal modes, in a Pr$^{\text{3+}}$ doped waveguide has been demonstrated \cite{seri2019quantum}. 

In rare-earth ions, entanglement generation between remote AFC-based memories has also been reported \cite{lago2021telecom, liu2021heralded}. In the former paper, a multimode AFC memory with 62 temporal modes has been used.


\onecolumngrid
\begin{center}
\begin{table}[h]
\scriptsize
\begin{tabular}{l c c c c c c}
   &\cellcolor{magenta}Quantum dots& \cellcolor{cyan}Diamond defect&\cellcolor{yellow}RE crystals &  \cellcolor{blue}~~~~Trapped ions~~~~& \cellcolor{green}Laser-cooled atoms& \cellcolor{red}Hot vapor   \\
  \hline
  \hline
  $\mathcal{F}$ or \color{blue}$\mathcal{C}\color{black}$ & $0.55^a$, $0.6^b$ & $0.54^c$ & $0.92^d$ , $0.8^e$ & \makecell{$0.58(0.88)^f$, $0.86^g$}&\makecell{ $0.72(\color{blue}0.38\color{black})^h$, $0.62^i$}& \color{blue}$\sim0.004$\color{black}~$^j$ \\
  \hline
  $d (m)$ & $5^a$, $2^b$& $30(2)^c$ & $50^d$ , $10^e$ & $520^f$, 50k$^g$& 22k(50k)$^h$, 33k$^i$& 0.3$^j$ \\
  \hline
  $p(\%)$ & $\sim7^a$, $\sim7^b$ & 5$^c$  & $\sim0.007^e$ & $\sim70^f$, 50$^g$& 3.8(1.5)$^h$, $\sim$1$^i$ & $\sim 0.04^j$ \\
  \hline
  $R_e$(Hz) & 2.3k$^a$, 7.3k$^b$ & \color{blue}$\sim$0.01\color{black}$^c$ & 1.4k$^d$,\color{blue}$\sim$0.3m\color{black}$^e$&\makecell{\color{blue}$\sim$0.5($\sim$0.06)\color{black}$^f$, $>1 (est.)^g$}& \makecell{\color{blue}$\sim$7m\color{black}($\sim$1.54)$^h$, \color{blue}$\sim$0.01\color{black}$^i$}& $<1 (est.)^j$ \\
  \hline
  $\lambda$(nm) & $953^a$, $968^b$ & $637^c$ & \makecell{$606\&1436^d$, $880^e$} & \makecell{$854^f$, $854\&1550^g$}& \makecell{$795\&1342^h$, $780\&1517^i$} & $850^j$ \\
  \hline
\end{tabular}
\caption
{Summary of the recent demonstrations of entanglement distribution using atomic memories highlighting measured fidelity ($\mathcal{F}$) or concurrence ($\mathcal{C}$), link distance ($d$), entangled-pair probability ($p$), entanglement rate ($R_e$) and wavelength ($\lambda$). Corresponding reference as a: \cite{delteil2016generation}, b: \cite{stockill2017phase}, c: \cite{pompili2021realization}, d: \cite{lago2021telecom} (Light-matter entanglement with RE crystal over a metropolitan network of field-deployed fibers has been reported in a recent paper \cite{rakonjac2023transmission}.), e: \cite{liu2021heralded}, f: \cite{krutyanskiy2023entanglement}, g: \cite{krutyanskiy2019light}, h: \cite{yu2020entanglement}, i: \cite{van2022entangling} and j: \cite{li2021heralding}.  Estimated values (est.) are authors' estimations based on reported quantities in corresponding references. Entanglement rates ($R_e$) shown in blue were achieved for two-photon or three-photon interference. }
	\label{fig:table_QN}
\end{table}
\end{center}
\twocolumngrid

\subsection{Advances in emissive quantum memories}

DLCZ-type memories can be used to generate a collective spin excitation in an ensemble of atoms and then recall it on demand as a propagating photon. A DLCZ-based quantum memory with telecom-wavelength conversion and a storage time of 100 milliseconds has been reported in cold Rb atoms \cite{radnaev2010quantum}. In addition, by confining the same platform inside a ring cavity, to enhance atom-light coupling, a DLCZ memory with a recall efficiency of $76\%$ and a sub-second lifetime has been demonstrated \cite{yang2016efficient}. More recently, $38\%$ memory efficiency and 0.92 fidelity for 0.1s storage has been achieved \cite{wang2021cavity}.

Entanglement generations between remote emissive LAMs have been demonstrated using different platforms, such as NV centers \cite{humphreys2018deterministic}, hot vapors \cite{li2021heralding}, trapped neutral atoms \cite{hofmann2012heralded}, trapped ions \cite{krutyanskiy2023entanglement,hannegan2022entanglement}, atomic ensembles \cite{yuan2008experimental}, and quantum dots \cite{delteil2016generation, stockill2017phase}. In particular, entanglement has been established between three distant quantum memories, each of which is made up of a laser-cooled Rb atomic ensemble \cite{jing2019entanglement}. 
Later, the same group reported entanglement generation between two laser-cooled atomic memories separated by  {22 km (field-deployed fibers) and 50 km (coiled fibers)} using a two-photon \cite{simon2003robust}, and a single-photon interference scheme \cite{duan2001long}, respectively \cite{yu2020entanglement}. Entanglement of two single trapped-atom over 33km telecom fiber has been achieved \cite{van2022entangling}. More recently, entanglement distribution  {over 50 km-long fiber spools} has been demonstrated using trapped ions \cite{krutyanskiy2022telecom}. In this experiment, ion-photon entanglement is first generated via cavity-mediated Raman transition \cite{schupp2021interface}. The generated photons in a middle station are then sent over 25 km of fiber in opposite directions. A deterministic Bell state measurement between nearby ions  then projects the state of the remote photons into an entangled state. A multi-node quantum network has been realized with NV centers in diamond \cite{pompili2021realization}.

There have been more recent proposals for designing quantum repeaters based on emissive memories \cite{asadi2020protocols, sharman2021quantum}.
Multiplexed memories can help with the entanglement generation rate of a repeater protocol \cite{sangouard2011quantum, asadi2018quantum}. In this regard, a spatially multiplexed DLCZ-type memory including 225 individually addressable elements has been demonstrated in a cold Rb atomic ensemble \cite{pu2017experimental}. Besides, although DLCZ-type memories are not ideal for temporal multiplexing, by combining this memory protocol with an AFC-based rephasing mechanism, a multimode memory with 12 and 11 temporal modes has been demonstrated in an ensemble of Eu$^{\text{3+}}$:YSO and Pr$^{\text{3+}}$:YSO, respectively \cite{laplane2017multimode, kutluer2017solid}. More recently, a multiplexed DLCZ memory with up to 10 temporal modes has been demonstrated in cold Rb atoms \cite{heller2020cold}. In this experiment, significant noise suppression has been achieved by using a CRIB-based rephasing mechanism and embedding the storage medium in a cavity.

 Long-distance entanglement generated between emissive memories has been used to conduct a loophole-free Bell test \cite{hensen2015loophole,rosenfeld2017event}. 

\subsection{Advances in nonlinear atomic memories}
Strong photon-atom interaction inside high cooperativity cavities can lead to near-deterministic photon-atom entanglement generation \cite{duan2004scalable,krutyanskiy2019light, yu2020entanglement}. In 2007,  reversible state transfer between light and a single atom trapped in a Fabry-Perot cavity has been realized \cite{boozer2007reversible}. Later, with a memory efficiency of $9.3\%$, Raman storage of laser pulses using single Rb atoms trapped inside a single-sided cavity was demonstrated \cite{specht2011single}. The same group has also demonstrated the storage of single photons in single atoms \cite{ritter2012elementary}.  Also, Schrödinger-cat states of entangled atom-light have been deterministically created by cold atoms placed in an optical cavity \cite{hacker2019deterministic}. Relying on cooperative and directional scattering, the generation of strong photon-photon correlation has also been demonstrated in these systems \cite{thompson2006high}.

In solid-state platforms, a quantum dot inside a nano-beam cavity was used to demonstrate near-deterministic photon-photon entanglement \cite{sun2018single}, and asynchronous BSMs of two photons have been demonstrated with SiV center coupled to nanophotonic resonators \cite{bhaskar2020experimental}.

Moreover, as discussed in the context of Rydberg atoms, a collective Dicke state prepared in an atomic ensemble by the combined action of the excitation blockade and interaction-induced dephasing can be efficiently mapped into a propagating single-photon wave-packet \cite{dudin2012strongly}. If only part of the collective atomic excitation is mapped into a retrieved field, an entangled state is created between the atomic excitation and a propagating light field \cite{li2013entanglement, li2016quantum, sun2022deterministic}. An ensemble of atoms can also be used as a source of single photons. By storing a part of the single-photon wave-packet in a separate atomic ensemble containing several million atoms, atom-light entanglement can be achieved. Separating the generation of single photons and their storage will allow to both achieve single-photon generation and memory-light mapping efficiencies in excess of $80\%$ \cite{wang2019efficient, ornelas2020demand} and memory storage times in excess of ten seconds \cite{dudin2013light}.

\subsection{Advances in QM materials and multiplexing techniques}

Using cold atoms long-life quantum storage (up to a record 16 seconds) \cite{dudin2013light}, multiplexed quantum storage, memory enhanced entanglement distribution scaling \cite{pu2021experimental} has been demonstrated. Spatial multiplexing can be achieved in cold-atom systems by segmenting a dense ensemble of cold atoms \cite{lan2009multiplexed, pu2017experimental} and temporal multiplexing by storing multiple temporally distinct modes \cite{heller2020cold} as high as memory’s delay-bandwidth product. More than one hour of coherence time has been achieved in trapped ions \cite{wang2021single}. Storage time of 10s has been demonstrated in a dual-species trapped-ion system, in which $^{\text{88}}$Sr$^{\text{+}}$ ion acts as communication qubit and $^{\text{43}}$Ca$^{\text{+}}$ ion serves as memory qubit \cite{drmota2023robust}. Quantum memory with a chain of 218 trapped ions has been demonstrated \cite{yao2022experimental}.  In addition, spatial multiplexing of laser-cooled atomic memories has been achieved that enabled the creation of up to 225 independent memory cells \cite{lan2009multiplexed, pu2017experimental, chang2019long}. Wave-vector multiplexing is another approach for multimode storage in a single atomic cloud with feed-forward control feasibility \cite{parniak2017wavevector, tian2017spatial}.

Defect centers and rare-earth ions in crystalline hosts are being considered suitable platforms for integrated quantum photonic applications. For example, silicon vacancy in 1D photonic crystals was recently used as memory qubits to improve the quantum key distribution rate \cite{bhaskar2020experimental} and rare-earth ions in yttrium orthosilicate and orthovanadate crystals were used for quantum storage \cite{zhong2017nanophotonic, liu2020demand}, quantum transduction \cite{bartholomew2020chip} and spin-photon interfaces \cite{kindem2020control, chen2020parallel}. 
The main limitations of the solid-state approach are inhomogeneity in atom/defect properties (such as transition frequency, dipole orientation, location, etc.) and decoherence due to noisy host materials especially when it comes to photonic integration where the emitters are proximal to etched surfaces. A couple of defect systems have emerged in the past years which demonstrate superior properties in both of these regards. Divacancy-defects in SiC can operate as a high-quality spin-photon interface \cite{anderson2019electrical} with spin coherence times reaching 5 seconds \cite{anderson2022five}. Recently, T-centers in silicon shows significant promise with narrow optical linewidths in the telecom O-band and long-lived spins with millisecond coherence times \cite{higginbottom2022optical, higginbottom2022memory}. Some of the rare-earth ions being considered include Erbium (Er$^{\text{3+}}$) and Ytterbium (Yb$^{\text{3+}}$) ions in a host crystal with telecom or near-telecom transitions and the ability of microwave-optical transduction, respectively \cite{asadi2022proposal}. Some of the host materials considered include Yttrium orthosilicate (Y$_{\text{2}}$SiO$_{\text{5}}$), which has been proven to provide long nuclear coherence time,  and lithium niobate (LiNbO$_{\text{3}}$)/Lithium Niobate on insulator (LNOI)  which has multifunctional capabilities. More recently, CaWO$_{\text{4}}$ \cite{le2021twenty, ourari2023indistinguishable} have attracted attention due to their low nuclear-spin-induced decoherence. Y$_{\text{2}}$O$_{\text{3}}$  \cite{FukumoriPRB2020, rajh2022hyperfine, gupta2023robust} is another promising hosts showing both narrow optical linewidths and millisecond long spin coherence times. Following the general scalings of spin qubit defect coherence \cite{kanai2022generalized}, at least a few dozens of known host crystals can provide a magnetically quiet matrix that supports $>$ms coherence time for spin defects. Many of those are potentially excellent host materials for rare-earth ions, and preliminary optical spectroscopy on a subset of them with favorable point group symmetries has been reported \cite{phenicie2019narrow, StevensonPRB2022}.

Among solid-state quantum memory systems, rare-earth ions doped crystals are also uniquely identified with their multimode quantum storage capability using different degrees of freedom \cite{seri2019quantum, gundougan2012quantum, tang2015storage}. In particular, the relatively large inhomogeneously broadened absorption profile of these ions has made temporal multiplexing the most convenient approach
. In this regard, the storage of more than a thousand temporal modes using rare-earth ions has set a new record in temporal multiplexing \cite{businger2022non}. Multiplexed storage of spatial, temporal, and spectral modes in a rare-earth crystal has also been demonstrated for close to 2000 modes and a projected mode capacity of 153k \cite{yang2018multiplexed, wei2022storage}. The feed-forward control of spectral modes has also been shown in rare-earth crystals \cite{sinclair2014spectral}.


\section{Vision}

Quantum memories have experienced rapid and broad advances over the last two decades. Still, much room for further developments remains. It is too early at this point to predict the physical platform which will offer the best performance for quantum memories. Given a wide array of material systems under consideration and an equally wide spectrum of e.m. fields to which these memories can be coupled, there may ultimately not be a single winner, but rather a range of quantum memories that are optimal for particular applications. Below we sketch some of the interesting quantum memory implementations and indicate both the promise and the challenges inherent to them.

Quantum memories based on hot atomic vapors are appealing for deployment in the field due to their less stringent environmental requirements \cite{wang2022field}. While such quantum memories are often noisy at high optical densities, the noise can be suppressed by employing an optical cavity. Nuclear spins of the noble gas can reach hours-long coherence times\cite{katz2021coupling, katz2022optical, shaham2022strong}. While direct optical access in noble gases is often difficult, a binary mixture of alkali and noble-gas atoms can function as a quantum memory with both convenient and strong coupling to light and exceptionally long storage times, and is therefore attractive for quantum repeater implementations \cite{ji2022proposal}. 

Laser-cooled atomic systems offer excellent storage times, high efficiency and fidelity, and versatile control options. With expected further advances in photonic integration and on-chip atomic traps \cite{chang2018colloquium, romaszko2020engineering, zhou2023coupling}, these systems might present sufficiently robust performance in compact packages to be considered for field implementations. 

Solid-state systems such as diamond color centers, quantum dots, defects in SiC, rare-earth ions, etc., offer outstanding large-scale integration possibilities. In particular, rare-earth doped crystals possess seconds- to hours-long coherence times \cite{zhong2015optically,ranvcic2018coherence}, with the ability to store telecom-band photons. Large spectral bandwidths and availability of temporal multiplexing protocols are noteworthy strengths of these systems. The electronic spins of the nitrogen-vacancy centers in diamonds have been shown to have millisecond-scale coherence times even at room temperatures, while even longer, second-scale, coherence times are achievable for their nuclear spins \cite{maurer2012room}. Access to the latter requires mapping quantum states between the electron and nuclear spins. The possibility of designing a quantum repeater using room-temperature NV centers has been discussed in Ref. \cite{Ji2022proposalroom}.

Taking advantage of atom-atom or atom-photon interactions under conditions of high cooperativity is promising for achieving near-deterministic approaches for scalable entanglement distribution with quantum memories. Cavity-enhanced interaction of photons with single trapped atoms, ions, or solid-state quantum centers enables deterministic photon-atom entanglement generation \cite{krutyanskiy2023entanglement, wang2021cavity,bhaskar2020experimental, reiserer2014quantum}. Similar functionalities are offered by ensembles of atoms with strong Rydberg interactions \cite{li2013entanglement}. Extending these approaches to the generation of photonic tree cluster states might make it possible to achieve one-way quantum repeater implementation for entanglement distribution over long distances \cite{borregaard2020one, zhan2020deterministic}.

It is likely that integrated and compact laser-cooled atomic systems will be advanced in parallel with solid-state quantum memories, in order to realize highly multiplexed quantum state storage. Further refinements in adapting optical cavities and atom-atom interactions in these systems are expected to overcome fundamental limitations associated with probabilistic entanglement creation schemes. In the long run, novel solid-state QOM materials will be integrated with photonic circuitry to realize fully integrated and scalable QOMs.
We expect integrated QOM systems to initially operate at milli-Kelvin to Kelvin temperatures and perhaps high B-fields. Eventually, materials will be available to allow multiplexed operation at $>$1K temperatures (and low B-fields) for large-scale operations (e.g. via efficient interfacing with telecom photons). Moreover, the implementation of multifunctional devices \cite{craiciu2021multifunctional, stas2022robust} in solid-state systems integrating quantum state generation, gate operation, storage, and non-destructive detection can open up possibilities for compact and scalable system design \cite{awschalom2021development}.  

We also note that by using satellites, it is possible to establish a global quantum network that does not heavily rely on quantum repeater ideas. However, even in those settings, it may be advantageous to use quantum memories. For example, they would enable quantum communication between distant stations on earth \cite{simon2017towards} or quantum teleportation over space-like separated nodes \cite{gundougan2021topical}.


\section{Acknowledgments}
 Y.L. and M.H. acknowledge the support from U.S. Department of Energy, Office of Science, Office of Advanced Scientific Computing Research, through the Quantum Internet to Accelerate Scientific Discovery Program under Field Work Proposal 3ERKJ381.
  F.K.A. and C.S. acknowledge the High-Throughput Secure Networks challenge program of the National Research Council of Canada. 
  A.K. acknowledges the support of the AFOSR and the NSF.

\bibliographystyle{ieeetr}
\bibliography{sample}{}

\end{document}